\documentstyle[12pt]{article}
\begin{document}
\baselineskip=21pt
\begin{titlepage}
\rightline{Alberta Thy-32-92}
\rightline{UMN-TH-1109/92}
\rightline{December 1992}
\vskip .2in
\begin{center}
{\large{\bf Physical Properties of Four Dimensional Superstring Gravity Black
Hole Solutions}}
\end{center}
\vskip .1in
\begin{center}

Bruce A. Campbell and Nemanja Kaloper

\vskip.2in
{\it Department of Physics, University of Alberta}

{\it  Edmonton, Alberta, Canada T6G 2J1}
\vskip.2in

Richard Madden and Keith A. Olive

\vskip.2in
{\it School of Physics and Astronomy, University of Minnesota}

{\it  Minneapolis, Minnesota, 55455 USA}

\vskip .2in

\end{center}
\centerline{ {\bf Abstract} }
\baselineskip=18pt

\noindent
We consider the physical properties of four dimensional black hole solutions to
the effective action describing the low energy dynamics of the gravitational
sector of heterotic superstring theory. We compare the properties of the
external field strengths in the perturbative solution to the full $O(\alpha')$
string effective action equations, to those of exact solutions in a truncated
action for charged black holes, and to the Kerr-Newman family of solutions of
Einstein-Maxwell theory. We contrast the numerical results obtained in these
approaches, and discuss limitations of the analyses. Finally we discuss how the
new features of classical string gravity affect the standard tests of general
relativity.

\end{titlepage}

\baselineskip=21pt

\newpage
{\newcommand{\la}{\mbox{\raisebox{-.6ex}{$\stackrel{<}{\sim}$}}}
{\newcommand{\ga}{\mbox{\raisebox{-.6ex}{$\stackrel{>}{\sim}$}}}
{\bf Introduction}
\vskip.3in

       Superstrings \cite{strev} have emerged as candidates to provide a
unified, quantum, description of matter, gauge forces, and spacetime geometry.
As prospective ``theories of everything" they are economical in their
fundamental principles. At the present (first quantized) level of description,
a string theory is specified by the fields (coordinate and other) carried by
the string world sheet. For consistent first quantized propagation of a
(closed) string, the quantum theory of the string world-sheet must satisfy
reparametrization and Weyl invariance, and the particular type of string
(bosonic, heterotic, type II) is determined by the number of local chiral
supersymmetries relating the world sheet fields. In a (super)conformal gauge
this requires that the world-sheet theory be a modular invariant conformal
field theory with total central charge (including gauge-fixing ghosts) equal to
zero. World sheet theories satisfying these conditions then define consistent
string propagation, and from the point of view of second-quantized string
theory would correspond to propagation in different classical solutions of the
second quantized theory.

       So the taxonomy of the string world sheet reflects the multiplicity of
classical solutions of the second-quantized theory. In this view, the number of
dimensions of spacetime, the gauge group, the fermion families, et cetera,
reflect the properties of a particular solution of the underlying theory of
second-quantized string. At the present time literally millions of
constructions for consistent string world-sheet theories are known, and we have
at present no method of classifying all solutions, or even of knowing if such a
classification exists. In the absence of a classification of string solutions,
we will only be able to extract predictions from, and make tests of, string
theory, if we can find general features of the plethora of string solutions
which must hold in any solution purporting to describe our universe.

       In fact, there is one universal feature shared by critical strings
propagating in spacetimes which are large compared to the characteristic string
size (strictly speaking, from the world-sheet point of view we should say
strings with world-sheet fields which include coordinate fields parametrizing a
manifold which is regular and large in string units). This is the presence in
the spectrum of closed strings of a tensor excitation, the metric (graviton);
the dynamics of this excitation computed from correlation functions of its
vertex operator reduce at long wavelengths (on the coordinate manifold) to the
dynamics of Einstein gravity (plus higher order corrections) on the manifold,
with this mode as the Einstein metric. In short, any critical string theory
solution with a large spacetime, predicts gravitation in that spacetime.
Historically it was this unexpected result that elevated string theory to
candidacy for a unified theory of all the interactions. In the present context,
the existence of gravitation in all string solutions with large spacetimes,
opens the prospect of direct tests of string theory by investigation of the
nature of string gravity, which is an inescapable consequence of the theory.

       Since it is the gravitational sector for which string theory gives
construction (solution) independent features, to test it one should consider
systems whose structure is essentially determined by gravitation. The unique
class of objects for which this is known to be true are the black holes that
arise in the theory. This is precisely the reason for the extensive studies of
the black holes in string theory conducted recently, where they have been
investigated with varying degree of approximation \cite{CMP} -
 \cite{kal4}. It is the purpose of this paper to analyze the
structure and physical properties of the classical black hole solutions that
arise in string gravity. We include the complete set of next order corrections
in $\alpha'$ (the inverse string tension) in the string gravity effective
Lagrangian, and examine the modifications that they produce to black hole
solutions both analytically and numerically. We will compare our results to
those that have been obtained when only a subset of the corrections are
included. We also examine how our solutions, considered as approximate exterior
solutions to astrophysical bodies, would modify the standard experimental tests
of relativistic gravity.


       In closed, oriented, critical, strings, the graviton excitation is part
of the gravitational multiplet of (classically) massless particles. It includes
the symmetric tensor graviton $g_{MN}$, a scalar dilaton $\phi$, and an
antisymmetric tensor two-form field $B_{MN}$. The effective action for the
leading order (independent of  $\alpha'$) gravitational dynamics of the long
wavelength modes (in the Einstein conformal frame) consists of the the Einstein
action for the graviton, a dilaton kinetic term, and a dilaton rescaled kinetic
term for the Kalb-Ramond form field. At the next order $O(\alpha')$ in the
expansion of the effective action for the heterotic string, there appear gauge
field kinetic terms for any gauge groups appearing in the string construction,
as well as higher order gravitational curvature terms, and Chern-Simons
modifications of the Kalb-Ramond field (plus of course construction dependent
matter sector fields which are not of interest to us). To $O(\alpha')$ the
effective Lagrangian describing the dynamics of the modes of interest to us may
be written \cite{2,3}:
\begin{equation}
S = \int d^4x \sqrt{g} \Bigl\{\frac{R}{2 \kappa^2} - 6 e^{-2 \sqrt{2}\kappa
\phi}  H^{\mu\nu\lambda}H_{\mu\nu\lambda} - \frac{1}{2} \partial_{\mu} \phi
\partial^{\mu} \phi + \frac{\alpha'}{16 \kappa^2} e^{-\sqrt{2}\kappa\phi}(\hat
R^2-F_{\mu\nu}F^{\mu\nu})\Bigr\}
\label{1}
\end{equation}
where the effective action (\ref{1}) has been expanded in the string tension to
$O(\alpha')$. $H_{\mu\nu\lambda}$ are the components of the three-form
\begin{equation}
H_{\mu\nu\lambda} = \partial_{[\lambda}B_{\mu\nu]} +
\frac{\alpha'}{8\kappa}(\Omega_{3L_{\mu\nu\lambda}}
-\Omega_{3Y_{\mu\nu\lambda}})\label{2}
\end{equation}
where $B_{\mu\nu}$ is the antisymmetric field in the gravitational multiplet,
and the gauge, and Lorentz, Chern-Simons terms are given by:
\begin{equation}
\Omega_{3Y_{\mu\nu\lambda}}=\frac{1}{2}
Tr(A_{[\lambda}F_{\mu\nu]}-\frac{2}{3}A_{[\mu}A_{[\nu}A_{\lambda]]})
\label{3}
\end{equation}
\begin{equation}
\Omega_{3L_{\mu\nu\lambda}}=\frac{1}{2}
Tr(\omega_{[\lambda}R_{\mu\nu]}
-\frac{2}{3}\omega_{[\mu}\omega_{[\nu}\omega_{\lambda]]})
\label{4}
\end{equation}
where the trace is taken over the (suppressed) gauge and Lorentz indices, and
the spin connection $\omega^{ab}_{\mu}=-e^{b\alpha}e^{a}_{\alpha;\mu}$ with
vierbein $e^{a}_{\mu}$. Note that the $H^2$ term in (\ref{1}) already contains
terms of $O(\alpha')$. In the action (\ref{1}) $\phi$ is the dilaton and the
Gauss-Bonnet combination is given by $\hat
R^2=R_{abcd}R^{abcd}-4R_{ab}R^{ab}+R^2$.

It is useful to dualize the Kalb-Ramond field in order to express the action in
terms of dynamical modes. Let us  re-express the action in terms of the dual of
$H$, $Y={^{\ast}} H$.
\begin{equation}
Y_{\sigma} = \sqrt{g} \epsilon_{\mu\nu\rho\sigma} H^{\mu\nu\rho}
\label{5}
\end{equation}
from equation (\ref{2}), we have the Bianchi identity
\begin{equation}
dH = \frac{\alpha'}{8\kappa}Tr(R \wedge R - F \wedge F)
\label{6}
\end{equation}
which we can write as
\begin{equation}
(\sqrt{g} Y^{\rho})_{,\rho} = \frac{\alpha'}{8\kappa}
\frac{\epsilon^{\mu\nu\lambda\sigma}}{4} (R_{ab\mu\nu} R_{\lambda\sigma}{}^{ba}
+ F_{\mu\nu} F_{\lambda\sigma})
\label{7}
\end{equation}
Using the constraint (\ref{7}), we can write the action in terms of the truly
dynamical axion mode of the theory as
\begin{eqnarray*}
S = \int d^4x \sqrt{g} \Bigl\{\frac{R}{2 \kappa^2} &-& \frac{1}{144} e^{2
\sqrt{2}\kappa \phi} \partial_{\mu} b \partial^{\mu} b  - \frac{1}{2}
\partial_{\mu} \phi \partial^{\mu} \phi \\
&+& \frac{\alpha'}{16 \kappa^2} e^{-\sqrt{2}\kappa\phi}(\hat
R^2-F_{\mu\nu}F^{\mu\nu})
 - b\frac{\alpha'}{8
\kappa}\frac{\epsilon^{\mu\nu\rho\lambda}}{4!\sqrt{g}}(R_{ab\mu\nu}
R_{\rho\lambda}{}^{ba} + F_{\mu\nu} F_{\rho\lambda})\Bigr\}
\end{eqnarray*}
\begin{equation}
\label{8}
\end{equation}
where $Y^{\rho}=\frac{1}{12} \exp(2\sqrt{2}\kappa\phi)(\partial^{\rho}b)$. If
one neglects the space-time dependence of the dilaton (consistently in the
perturbative expansion when dilaton dependence occurs at $O(\alpha')$), this
leads to the axion equation of motion
\begin{equation}
\Box a =\frac{6\alpha'}{8 \kappa}
\frac{\sqrt{2}\epsilon^{\mu\nu\rho\lambda}}{4!\sqrt{g}}(R_{ab\mu\nu}
R_{\rho\lambda}{}^{ba} + F_{\mu\nu} F_{\rho\lambda})
\label{9}
\end{equation}
where $a$ is the rescaled axion which approximates the one with canonical
kinetic term,
$\partial^{\rho}a=\exp(2\sqrt{2}\kappa\phi)\partial^{\rho}b/{6\sqrt{2}}$.
[Because we are anticipating that the dilaton $\phi \sim \alpha'$, we neglect
the exponentials in writing equation (\ref{9})]. The dilaton equation of motion
is now rather easily obtained from the action (\ref{8}), and the axion field
redefinition
\begin{equation}
\Box \phi =\frac{\alpha\acute{ }\sqrt{2}}{16
\kappa}\exp(-\sqrt{2}\kappa\phi)(\hat R^2-F_{\mu\nu}F^{\mu\nu}) +
\sqrt{2}\kappa  \exp(-2\sqrt{2}\kappa\phi)(\partial_{\mu} a)(\partial^{\mu} a)
\label{10}
\end{equation}
Again when the solution for $a$ is already of order $\alpha'$, we may drop the
last term (consistently in our expansion); expanding the exponential in the
dilaton as before, and keeping terms of order $\alpha'$, we arrive at
\begin{equation}
\Box \phi =\frac{\alpha'\sqrt{2}}{16 \kappa}  (\hat
R^2-F_{\mu\nu}F^{\mu\nu})
\label{11}
\end{equation}
Note that both the axion and the dilaton are first non-trivial at order
$\alpha'$.

In \cite{CDKO1,CDKO2,CKO1,CKO2} we have evaluated the source terms for the
axion and dilaton for the background of a weakly-charged, rotating black hole.
We convoluted these sources with the Green function for that background, to
compute the classical hair produced in the Kerr-Newman background. We kept all
effects to $O(\alpha')$ perturbatively for the slowly-rotating, weakly-charged
case. To the order in which we worked, back-reaction on the metric could be
neglected, and it remains of Kerr-Newman form [see appendix]. The explicit
forms of the dilaton and axion hair we obtained are also shown in the appendix.
It is our purpose in this paper to compare our perturbative solution to exact
solutions of truncations of our action, both perturbatively and numerically, as
well as to discuss the physically measurable effects of the dilaton and axion
hair produced.




\vskip.3in
{\bf Black Holes And String Theory}
\vskip.3in

In order to elaborate on the difference between the classical General
Relativity Kerr-Newmann family of black hole
solutions (KN),  our $\alpha'$ string dilaton/axion hair solutions \cite{CKO2}
(hereafter CKO) and the dilaton/axion solutions
by Gibbons \cite{Gibb}, Gibbons and Maeda \cite{GM}, Garfinkle, Horowitz and
Strominger \cite{GHS}, Shapere, Trivedi and Wilczek \cite{STW} and Sen
\cite{Sen} (which hereafter we refer to as GS), it will be useful to first
describe how these solutions are similar. For simplicity, we will consider the
non-rotating case.
The similarities and differences are not qualitatively affected when $A \ne 0$.
Recall that the solutions of  GS were found by beginning with the string action
(1) in the $\alpha'$ expansion, truncating the action to $O(\alpha')$ but
keeping only one of the many $O(\alpha')$ contributions, namely
the gauge kinetic term. With this truncation, they find the remarkable closed
form  solution given in the appendix.

To compare the GS solution (with $A = 0$) for the metric and the standard RN
solution we need to perform a coordinate transformation:
\begin{equation}
r^{2} = \rho^{2} - \frac{Q'^{2} \rho}{GM} =
{\tilde \rho}^{2} + \frac{{\tilde Q}^{2}{\tilde \rho}}{GM}
\label{12}
\end{equation}
\noindent This brings the angular part of the metric (the coefficient of
$d\Omega$) to the standard form. Solving for $\rho(r)$ and $\tilde \rho(r)$ and
keeping only terms of $O(\alpha')$ (note that
$Q'^2 = \tilde Q^2 = \alpha' Q^2 /8$ as described in the appendix) one sees
immediately that both metrics (\ref{A10}) and (A11) reduce exactly to the
Reisnner-Nordstrom metric in (A8). This should not be a surprise since it has
been shown that the Kerr-Newmann family is the $O(\alpha')$-expanded metric
solution in string theory \cite{CKO2}.

In this sense, the GS solution is in fact equivalent to the KN solution up to
an $O(\alpha')$  coordinate transformation, and hence equivalent to the metric
in the CKO solution. Similarly one can see that the dilaton and axion solutions
are also equivalent to $O(\alpha')$. For example, the dilaton solution of GS is
\begin{equation}
\tilde \phi = - \log{(1 + \frac{\tilde Q^2}{GM \tilde \rho})} \approx
- \frac{ \alpha' Q^2}{8GMr}
\label{13}
\end{equation}
\noindent which corresponds to (A12) given in the appendix.

However there are several important aspects to which these solutions differ.
First we compare the horizons and critical limits of these solutions (at least
formally) to perform the $\alpha'$ expansion with large $q^2$, in this case
$Q^2$ must be very large.
In the RN solution, the outer horizon occurs at the radial coordinate value
\begin{equation}
r_H = GM + \sqrt{G^2M^2 - Gq^2}
\label{14}
\end{equation}
\noindent and the critical limit (for which there is a naked singularity) is
obtained as
\begin{equation}
\alpha' \frac{Q^2}{8} = Gq^2 = G^2M^2 = Q'^2
\label{15}
\end{equation}

\noindent In the $\alpha'$ expansion of string
theory, Eq. (\ref{14}) can be expanded so that
\begin{equation}
r_H = 2GM - \frac{q^2}{2M} = 2GM - \alpha' \frac{Q^2}{16GM} = 2GM -
\frac{Q'^2}{2GM}
\label{16}
\end{equation}
\noindent Otherwise, it is no longer possible to obtain the critical limit from
this expression. In the G-S solution the horizon occurs at
\begin{equation}
\rho_H = 2GM
\label{17}
\end{equation}
\noindent which when related to $r$ by the coordinate transformation
Eq. (\ref{12}) becomes
\begin{equation}
\rho_H = r_H + \frac{Q'^2}{2GM} = 2GM
\label{18}
\end{equation}
\noindent so that $r_H$ is given precisely by (\ref{16}).(In terms of $\tilde
\rho$, $\tilde \rho_H = 2GM - \tilde Q^2 /M$ with $A = 0$.) The horizons in the
two solutions are again equivalent in $\alpha'$ expansion up to a coordinate
transformation. The critical limit is however different and now occurs when
\begin{equation}
Q'^2 = 2G^2M^2
\label{19}
\end{equation}
\noindent Note that despite $Q'^2 \propto \alpha'$, $Q'^2$ can be thought of as
large, so long as $Q^2$ is large, i.e. one can work in the limit of small
$\alpha'$ and fixed $Q'^2$. Similarly, the radii at which the surface area goes
to zero are also the same for small $\alpha'Q^2$.
In the GS solution
\begin{equation}
A = 4\pi \rho (\rho - \frac{Q'^2}{GM})
\label{20}
\end{equation}
\noindent the area vanishes at the hypersurface corresponding to $\rho = Q'^2 /
GM$. Under the coordinate transformation (\ref{12}),
\begin{equation}
A = 4\pi r^2
\label{21}
\end{equation}
\noindent as in the RN solution.

The temperature of the black holes also agree at small $Q'^2$ while they differ
at large $Q'^2$. Writing $T = K / 2\pi$ with
\begin{equation}
K = \frac{1}{2} \Bigl(\frac{\partial g_{00}}{\partial r}\Bigr)_{r=r_H}
\label{22}
\end{equation}
\noindent In the GS solution
\begin{equation}
K = \frac{1}{2} \Bigl(\frac{2GM}{\rho_{H}^{2}}\Bigr) = \frac{1}{4GM}
\label{23}
\end{equation}

\noindent For the RN solution,
\begin{equation}
K = \frac{\sqrt{G^2M^2 - Gq^2}}{G^2M^2(1 + \sqrt{1 - \frac{q^2}{GM^2}})^2}
\approx \frac{1}{4GM}\Bigl(1 - \frac{q^2}{2GM^2}\Bigr)\Bigl(1 +
\frac{q^2}{2GM^2}\Bigr) \approx \frac{1}{4GM}
\label{24}
\end{equation}
\noindent as in (\ref{23}). In the small $Q'^2$ limit, these expressions for T
are indistinguishable, while for large $Q'^2$, the difference is clearly
manifest.

In the large $Q'^2$ limit, the coordinate transformations (\ref{16}) are no
longer valid, and the solutions appear different. Thus it is important to
better understand this limit to determine whether or not these new solutions
are in some sense better stringy black hole solutions. To determine precisely
what that limit is, we need to inspect terms in the action which have been
discarded. Indeed, it is not obvious that higher order curvature terms such as
Gauss-Bonnet and
$\alpha' R^3$ \cite{2,3,BBJ}
terms are small compared with the terms kept in ($A1,A4$). The reason for
possible concern is that as one approaches the critical limit
$Q'^2 = 2GM^2$, the curvature scalar blows up on the horizon at $\rho = 2M$,
\begin{equation}
R = \frac{Q'^4 (2GM - \rho)}{2 \rho^3 (Q'^2 - GM\rho)^2} \rightarrow
\frac{8G^2M^2}{\rho^3(2GM - \rho)} \propto \frac{1}{x}
\label{25}
\end{equation}
\noindent with $x = 2GM - \rho$. Indeed, in the critical limit, other terms in
the action also blow up, such as
\begin{equation}
\partial_{\mu} \phi' \partial^{\mu} \phi' = \Bigl(\frac{Q'^2}{2(GM\rho -
Q'^2)}\Bigr)^2 \Bigl(1 - \frac{2GM}{\rho}\Bigr) \rightarrow
\frac{G^2M^2}{\rho^3(\rho - 2GM)} \propto \frac{1}{x}
\label{26}
\end{equation}
\noindent and
\begin{equation}
e^{-2\phi'} F^2 = \frac{Q'^2}{\rho^4} \Bigl(1 - \frac{2GM}{\rho}\Bigr)^{-1}
\rightarrow \frac{2G^2M^2}{\rho^3(2GM - \rho)} \propto \frac{1}{x}
\label{27}
\end{equation}
\noindent (here we give the expression for the electrically charged case, the
general case is equally divergent). One might expect a priori, that other
$O(\alpha')$ terms would blow up more strongly (as $x^{-2}$ or worse). Should
this occur, the truncation keeping only the $\alpha' F^2 \exp{(-2\phi')}$ term
would be invalid except at small $Q'^2$ where all solutions are equivalent
anyway. As we will see, the answer to this question depends heavily on whether
or not the black hole posseses any \underbar{magnetic} charge.

First, let us recall the generalization of these solutions when both electric
and magnetic charges are present. In the KN solution $q^2 = q^{2}_{e} +
q^{2}_{m}$ and in the CKO solution for the dilaton (and axion) the general
solution is given in ($A12,A13$). The GS dilaton solution can be generalized by
taking $Q'^2 = Q'^{2}_{e} + Q'^{2}_{m}$ and (\cite{STW})
\begin{equation}
e^{-2\phi'} = Q'^2 \frac{1 - \frac{Q'^2}{GM\rho}}{Q'^{2}_{m} + Q'^{2}_{e}(1 -
\frac{Q'^2}{GM\rho})^{2}}
\label{28}
\end{equation}
\noindent or in the critical limit
\begin{equation}
e^{-2\phi'} = \frac{2G^2M^2 \rho (\rho - 2GM)}{Q'^{2}_{m}\rho^2 +
Q'^{2}_{e}(\rho - 2GM)^{2}} = - \frac{2G^2M^2 \rho x}{Q'^{2}_{m}\rho^2 +
Q'^{2}_{e}x^2}
\label{29}
\end{equation}
\noindent As one can see, the limiting behavior of $\exp{(-2\phi')}$ as $x
\rightarrow 0$ depends crucially on whether or not $Q'_{m} = 0$.

To test the limit, we examine several additional $O(\alpha')$ contributions.
For example, the Gauss-Bonnet combination (in the critical limit)
\begin{equation}
\alpha' e^{-2\phi'} \hat R^2 = \alpha' \frac{16G^2M^2(2GM - 3\rho)}{\rho^6
(\rho - 2GM)} e^{-2\phi'} = \alpha' \frac{32G^4M^4}{\rho^5} \frac{(2GM -
3\rho)}{Q'^{2}_{m}\rho^2 + Q'^{2}_{e}x^2}
\label{30}
\end{equation}
\noindent which for $Q'_{m} \ne 0$ is {\it finite} as $x \rightarrow 0$ but
diverges as $x^{-2}$ if $Q'_{m} = 0$. There are additional dilaton terms
\begin{equation}
\alpha' e^{-2\phi'} (D^{\mu} \partial^{\nu} \phi')(\partial_{\mu} \phi')
(\partial_{\nu} \phi') \rightarrow \cases{~~\alpha' \frac{G^3M^3(3GM -
2\rho)}{\rho^7 x} ~~~&($Q_e=0$) \cr
- \alpha' \frac{G^3M^3(3GM - 2\rho)}{\rho^5 x^3} ~~~&($Q_m=0$) \cr}
\label{31}
\end{equation}
\noindent and
\begin{equation}
\alpha' e^{-2\phi'} \Bigl((\partial_{\mu} \phi') (\partial^{\mu} \phi')\Bigr)^2
\rightarrow  \cases{- \alpha' \frac{G^4M^4}{\rho^7 x} ~~~&($Q_e=0$) \cr
- \alpha' \frac{G^4M^4}{\rho^5 x^3} ~~~&($Q_m=0$) \cr}
\label{32}
\end{equation}
\noindent Although the $Q_e = 0$ contributions diverge as $x^{-1}$ (as do the
terms included in ($A1,A4$)), they are suppresed by a factor $\alpha'G^2M^2 /
\rho^4 \propto \alpha' / G^2M^2 \propto 1 / GM^2$. Then for massive black
holes, $GM^2 >> 1$, even these divergent terms can be safely neglected. However
for $Q_m = 0$, these terms diverge as $x^{-3}$ and can not be neglected when
studying the critical behavior or the solution near the horizon. Similarly
other terms in the action to order $\alpha'$ involving only $\phi'$, $R$ or $F$
show the same behavior. They diverge as $x^{-1}$ but are suppresed by $1/GM^2$
when $Q_m \ne 0$ and diverge as $x^{-3}$ when $Q_m = 0$. We also check the
model independent curvature cubed term at $O(\alpha'^2)$, \cite{BBJ}
\begin{equation}
\alpha'^2 e^{-4\phi'} \bigl(3I_1 - 4I_2 \bigr) \rightarrow 48 G^3M^3\alpha'^2
\frac{13G^3M^3 - 12G^2M^2 \rho + 4GM \rho^2 -2 \rho^3}{\rho^{11}}
\cases{\frac{1}{x} ~~~&($Q_e=0$) \cr
\frac{\rho^4}{x^5} ~~~&($Q_m=0$) \cr}
\label{33}
\end{equation}
\noindent which again only diverges as $x^{-1}$,
and is suppresed as $\alpha'^2 / G^4M^4 \propto 1/ G^2M^4$
when $Q_e = 0$ but severely
diverges as $x^{-5}$ when $Q_m = 0$. The curvature cubed
invariants $I_1, I_2$ are defined by$^{\cite{BBJ}}$
\begin{eqnarray}
& & I_1 = R_{\mu\nu}{}^{\alpha\beta}~R_{\alpha\beta}{}^{\gamma\sigma}
{}~R_{\gamma\sigma}{}^{\mu\nu} \nonumber \\
& & I_2 = R_{\mu\nu}{}^{\alpha\beta}~R_{\nu}{}^{\gamma\beta\sigma}
{}~R_{\gamma\mu\sigma}{}^{\alpha}
\label{33a}
\end{eqnarray}

To understand the pathological behavior of the solutions in the critical limit
we can restrict our attention to purely electric or magnetic solutions. Then,
as was described in \cite{GHS} , the gauge fields in the two cases are related
by a duality transformation involving the dilaton, itself transforming
according to $\phi \rightarrow - \phi$. It is this sign difference which is
responsible for the stability of the magnetically charged critical solutions
under $O(\alpha')$ corrections. Indeed, in this case we can conformally
transform the metric to the world-sheet frame according to $g_{\mu\nu} =
\exp(-\sqrt{2}\kappa\phi) G_{\mu\nu}$. The metric takes the form
\begin{equation}
ds^2_{WS} =  dt^2 - \bigl(1 - \frac{2GM}{\rho}\bigr)^{-2} d\rho^2 - \rho^2
d\Omega
\label{34}
\end{equation}
\noindent and as Garfinkle et al. \cite{GHS}
show is non-singular. In fact it corresponds
to the geometry of an infinite throat, with the singularity removed to
(unreachable) spatial negative infinity. As such, higher order curvature
corrections in the $\alpha'$ expansion may be expected to be benign for large
black holes, which have large ``diameter'' throat. Since the world-sheet action
then takes the form
\begin{equation}
S = \int d^4x \sqrt{G} e^{-\sqrt{2}\kappa\phi}\Bigl\{\frac{R_{WS}}{2 \kappa^2}
- 6  H^{\mu\nu\lambda}H_{\mu\nu\lambda} + \partial_{\mu} \phi
\partial^{\mu} \phi + \frac{\alpha'}{16 \kappa^2} (\hat
R^2_{WS}-F_{\mu\nu}F^{\mu\nu}) + \ldots \Bigr\}
\label{35}
\end{equation}
\noindent and all the terms in the parenthesis behave very softly inside the
throat (the most singular terms being the dilaton contributions), we see that
the worst singular behavior originates from the effective coupling
$\exp(-\sqrt{2}\kappa\phi)$ and its variation. In the electrically charged
critical solution the situation is reversed due to the fact that the dilaton
solution is the inverse of the magnetically charged one and hence the picture
of an infinite throat world-sheet geometry is destroyed, as conformal rescaling
makes singularities harder. Thence the difference between the $Q_e = 0$ and
$Q_m = 0$ cases. We should note, however, that although the magnetically
charged critical solution is stable under $O(\alpha')$ (and possibly higher)
corrections, it is subject to large quantum corrections; as is well known, the
quantum corrections in string theory correspond to higher genus world-sheets,
involving powers of the effective coupling $\exp(- \sqrt{2}\kappa\phi)$ (the
loop counting expansion parameter) in the effective action. Hence going deeper
into the throat in the magnetically charged critical case would make these
contributions more significant, eventually threatening to take us outside of
the perturbative regime of the theory \cite{STW}.

We note that the rotating solution of Sen is free of the pathology (the
appearance of corrections which are more singular than the terms in the
original $O(\alpha')$ action) discussed above due to the fact that the presence
of the angular momentum regularizes the solution in the critical case, much the
same way as the electric-magnetic solutions of $N = 4$ supergravity with two
independent $U(1)$ gauge field charges \cite{kal1} - \cite{kal4}. We recall
that the horizon of Sen's solution is
$\tilde \rho_H = GM - \tilde Q^2/2GM + \sqrt{(GM - \tilde Q^2/2GM)^2 - A^2}$,
so the critical limit occurs when $GM - \tilde Q^2/2GM = \mid A \mid$, implying
that $\tilde \rho_H = \mid A \mid > 0$ when $A \ne 0$, whereas the singularity
is located at $\tilde \rho_S = 0$ irregardless of the value of angular
momentum. The nonrotating GS solution then appears as a special case of the $N
= 4$ supergravity class. Observation of critical solutions in these two cases
demonstrates that even in the critical limit the singularity and the horizon
remain separated, except in the special case when the solutions reduce back to
GS, as can be seen from the fact that the area of the critical horizons remains
nonzero. We should note that in the context of supergravity and superstring
models a non-renormalization theorem has been proved \cite{kal1},\cite{kal2}
stating that the classical action for critical solutions is not changed by
quantum fluctuations derived from its path integral. The crucial point in the
proof is the presence of unbroken supersymmetries in the critical backgrounds.
However, in these cases metric in the world-sheet conformal frame is not
singularity-free and hence the solutions may be sensitive to inclusion of
higher order $\alpha'$ terms, corresponding to multi-string classical
scattering.

We can now conclude the following regarding the difference between the black
hole solutions. In the strict $\alpha'$ expansion (when $Q'^2$ is also small)
the GS and CKO solutions are equivalent in all respects up to a coordinate
transformation (for $GM^2 \gg 1$). The metric is then also the same as KN, as
the backreaction occurs first at $O(\alpha'^2)$. When $Q'^2$ is large and $Q_m
\ne 0$, and $GM^2 \gg 1$, the GS solution offers the best description of the
stringy black hole metric and dilaton/axion hair. The truncation used in
obtaining the solution is not valid in the critical limit when $Q_m = 0$. When
$Q'^2$ is small and $GM^2 ~\ga~ 1$, the CKO solutions for dilaton/axion hair
are
expected to be better, as the Gauss-Bonnet correction has been included.
Finally and unfortunately, when $Q'^2$ is large and $GM^2 \approx 1$, none of
these solutions is expected to give an accurate description of the stringy
black hole. In the next section, we make some numerical comparisons of the
dilaton and axion hair in the two solutions.



\vskip.3in
{\bf Numerical Comparisons Of Solutions}
\vskip.3in

The axion and dilaton solutions given in Eqs (A12) and (A13) represent
expansions (in the angular momentum) of the exact ($O(\alpha')$) integral
solutions
in terms of Green's functions. The exact equations were integrated numerically
and
compared (for the monopole, dipole and quadrupole moments)
with the expansions given in the appendix.
As expected for small values of A, the expansion is quite good.  Even
for values of $A/GM$ as large as 0.8, the difference between the exact solution
and the expansion was only 10\%.

We now compare the solution of CKO  with Sen's exact
solution to the truncated action with the coordinate change $r^2 \rightarrow
\tilde \rho^2 + \tilde Q^2 \tilde \rho /M$ reducing the angular part of the
metric to the standard
form. We concentrate on three limiting cases:
\begin{enumerate}
\item $G M^2 = 1$ and $Q'^2 \ll 1$
\item $G M^2 \gg 1$ and $Q'^2 / 2 \ll G M^2$
\item $G M^2 \gg 1$ and $Q'^2 / 2 \approx G M^2$
\end{enumerate}
In the fourth logical case, $G M^2 \approx 1$ and $Q'^2 / 2 \approx G M^2$
neither solution is expected to be good and no comparison was made.

As we indicated in the previous section, in case (1) we expect the solutions
to differ because of the inclusion of the Gauss-Bonnet corrections used by CKO.
In this regime the CKO solution is more accurate.
In Figure 1, where we chose $Q_e^2 = 1$ (or $Q'^2 = 1/8$ with $\alpha' = 1$)
 we see the results of this comparison for the axion (1a) and the dilaton
(1b). One can see a clear numerical difference for the two solutions due to the
importance of the Gauss-Bonnet contributions in this regime.
Case (2) is shown in Figure 2. Here we chose $G M^2 = 10^4$ and
$(Q'^2/G M^2) = 1.25 \times 10^{-3}$. Our expectation that the solutions are
equivalent
is clearly borne out.

 The final case was a nearly extremal
black hole. In this case we expected the GS solution to be better than CKO
since we are violating the condition that the charge be small. The results
shown in
Figure 3  for $G M^2 = 10^4$ and $(Q'^2/G M^2) = 1.53$,
unexpectedly show that the solutions remain close even as the extremal
limit is approached. However this similarity may be deceptive
since the background spacetimes of
GS and CKO differ considerably in this limit. In this particular case,
the horizon of the GS spacetime is located at an $r$ coordinate value
approximately half that of CKO and the field strength may be continued to much
larger values at this smaller r value. In general, as the extremal limit is
approached the horizon of GS retreats to a coordinate value of $r=0$ while
CKO remains fixed at $r=2M$ since CKO does not consider the effect of
back-reaction of the fields on the metric.

\vskip.3in
{\bf Motion Of Test Particles And Force From Hair}
\vskip.3in

In this section we will address the question of observing the field hair which
we discussed previously. We will attempt to compute measurable effects of the
hair upon probes moving in such backgrounds, and analyze deviations from
General Relativity that so arise. To gain insight concerning the nature of such
effects under the most general circumstances, we will discuss some limiting
situations and argue on their basis what happens in more general cases.

We will start by discussing the motion of probes of various structure (made of
gravitons, photons, axions, $\ldots$) and attempt to determine what is the
influence of the hair on them. For simplicity's sake, we will assume as our
first case that the black hole background to $O(\alpha')$ is Schwarzschild,
with dilaton hair arising from the Gauss-Bonnet combination. Then a quick
glance at the action supports one's naive expectation that the hair should
influence motion of different particles differently, for the fields appearing
in the action do not take simultaneously canonical form for their kinetic
terms. This is due to their coupling to the dilaton, induced on the
world-sheet, where all fields couple to it in the same way. Yet we argue that
in order to discuss the low energy limit of string theory we should conformally
transform the action to the Einstein frame, where the graviton is rendered
canonical. Then we can use the notions of standard GR to analyze the solutions
since the deviations can be interpreted as a consequence of the extra fields
appearing in the theory.

It was argued by Shapere, Trivedi and Wilczek \cite{STW} that the motion of
quanta of the dilaton and the axion can be determined by conformally rescaling
the metric so that in the resulting background the kinetic term of the
respective field is canonical and all explicit reference to the dilaton
coupling disappears. Then, by analogy with the photon in standard GR, the
particles are required to move along null geodesics of the conformal transform
of the metric. Moreover, as is well known, null geodesics do not change under
conformal transformations, and the only difference found in \cite{STW} in the
motion of different test particles was manifest in the affine parameter
measuring intrinsic geometry of a trajectory. As a consequence, it was
concluded that the dilaton effects amount to changes in the appearance of the
global structure of space-time to different observers as well as variations in
local physics, associated with observers in conformally related frames. This
argument, however, is not relevant to the motion of photons because the Maxwell
Lagrangian density $\sqrt{g} F^2$ is a conformal invariant in four dimensions.
In other words, no conformal rescaling can yield the canonical kinetic term for
the photon. Thus the issue of the motion of photons in a background with hair
was left unresolved as ever.

Here we propose an alternative approach to the derivation of the dilaton hair
induced force on test particles. We study the geometric optics approximation of
a corresponding field equation describing motion of a highly localized solitary
wave train. We invoke the validity of the Principle of Equivalence in the
Einstein frame, conduct our calculations in a flat Minkowski background, and
extend the results to a curved one via the standard minimal prescription,
following from the Principle of Equivalence. Our result agrees with the
ans\"atz of \cite{STW} in the case of the axion, and extends it to the photon,
giving a very similar formula. We then investigate the influence of this,
dilaton-induced, force on the motion of test particles.

To start with, we can write down the equations of motion for a generic
antisymmetric tensor field, resembling those which can be derived from the low
energy action for the axion and the photon. They are the Euler-Lagrange
equation,
\begin{equation}
\partial_{\mu} e^{- \xi \kappa \phi} F^{\mu\nu\lambda \ldots} = 0
\label{36}
\end{equation}
\noindent where $\xi$ is $\sqrt{2}$ for the photon and $2\sqrt{2}$ for the
axion, and the Bianchi identity following from the definition of the field
strength as an exact form:
\begin{equation}
\partial_{[\mu}  F_{\nu\lambda\sigma \ldots]} = 0
\label{37}
\end{equation}
\noindent Here we have defined the field strength $p$-form
$F_{\mu\nu\lambda\ldots}$  as the exterior derivative of a potential
$p$-$1$-form $A_{\nu\lambda\ldots}$, $~F_{\mu\nu\lambda\ldots} = p!
\partial_{[\mu} A_{\nu\lambda\ldots]}$. Since by its definition $F$ is
invariant under $U(1)$ gauge transformations, a gauge has to be fixed for
ambiguities to disappear. We choose to work in the Lorenz gauge:
$\partial_{\mu}  A^{\mu\nu \ldots} = 0$. As we are interested in the motion of
a solitary wave train, representing a particle, we use for the solution the
following ans\"atz:
\begin{equation}
A_{\mu\nu \ldots} = f_{a~ \mu\nu \ldots} e^{{\rm i} S}
\label{38}
\end{equation}
\noindent where $f_{a~ \mu\nu\ldots}$ is the (antisymmetric) polarization
tensor of the particle (wave train), and the phase $S$ is the classical action
of the particle (which solves the corresponding Hamilton-Jacobi equation). The
index $a$ counts all possible polarization states. The gauge condition then
reads $f_{a}{}^{\mu\nu \ldots} \partial_{\mu} S = 0$. Upon substitution of this
ans\"atz in the equations of motion, noting that $F_{\mu\nu\lambda\ldots} =
{\rm i} p! \partial_{[\mu}S f_{a~ \nu\lambda\ldots]} e^{{\rm i} S}$, we find
that the Bianchi identity is satisfied identically (as one expects) and that
the Euler-Lagrange equation yields, after canceling the phase $\exp({\rm i} S)$
and separating real and imaginary parts
\begin{eqnarray}
\partial_{\mu}S \partial^{[\mu}S f_{a}{}^{\nu\lambda\ldots]} = &0& \nonumber \\
\partial_{\mu} \Bigl( e^{- \xi \kappa \phi} \partial^{[\mu} S
f_{a}{}^{\nu\lambda\ldots]}\Bigr) &=& 0
\label{39}
\end{eqnarray}
\noindent We recall the orthogonality relations of the polarization tensors,
$f_{a}{}^{~\mu\nu\ldots}f_{b~ \mu\nu\ldots} = \delta_{a~b}$. This at hand,
after contracting the equations above with  $f_{b~ \mu\nu\ldots}$ we obtain
\begin{eqnarray}
\bigl(\partial S \bigr)^2 &\delta_{a~b}& - ~\partial_{\mu} S \partial^{\nu} S
f_{a}{}^{\mu\lambda\ldots}f_{b~ \nu\lambda\ldots} - ~ \bigl( ~{\rm other
{}~transpositions}~\bigr) = 0 \nonumber \\
\partial_{\mu} \Bigl( e^{- \xi \kappa \phi} \partial^{\mu} S
\Bigr)&\delta_{a~b}& - ~\partial_{\mu} \Bigl( e^{- \xi \kappa \phi}
\partial^{\nu} S \Bigr) f_{a}{}^{\mu\lambda\ldots}f_{b~ \nu\lambda\ldots} - ~
\bigl(~{\rm other ~transpositions}~\bigr) = 0
\label{40}
\end{eqnarray}
\noindent With use of the gauge condition $f_{a}{}^{\mu\nu \ldots}
\partial_{\mu} S = 0$ we see that all the terms but the first in these two
equations are zero identically. Thus the field equations reduce in this limit
to the particularly simple form
\begin{eqnarray}
& &~~~~~~~\bigl(\partial S \bigr)^2 = 0 \nonumber \\
& &\partial_{\mu} \Bigl( e^{- \xi \kappa \phi} \partial^{\mu} S \Bigr) = 0
\label{41}
\end{eqnarray}
\noindent These two equations constrain the classical action of the wave train.
With their help we will derive the action for the motion of the wave packet,
and compute the dilaton force acting on it. We note that the first of the two
equations is nothing else but the classical relativistic Hamilton-Jacobi
equation describing motion of free massless particles. This is almost exactly
what one should expect to be the case, considering that the particles are
associated with fields which are massless due to their gauge symmetry. We say
almost, since we do expect the dilaton to influence the motion, so it is not
that of free particles. To consider this, we recall that if we define the
vector field $V_{\mu}$ along the trajectory $\Gamma$ of a particle according to
\begin{equation}
V_{\mu} = \partial_{\mu} S\mid_{\Gamma}
\label{42}
\end{equation}
\noindent (where this definition is always possible at least locally) we can
rewrite the total action as the line integral
\begin{equation}
S = \int_{\Gamma} V_{\mu}~dx^{\mu}
\label{43}
\end{equation}
\noindent after eliminating the first derivatives with help of the equations of
motion. The simple substitution $dx^{\mu} = \dot x^{\mu} d\lambda$ valid along
the trajectory $\Gamma$ where $\lambda$ is the parameterization of $\Gamma$,
enables us to rewrite the action above as
\begin{equation}
S = \int_{\Gamma} V_{\mu} \dot x^{\mu} d\lambda
\label{44}
\end{equation}
\noindent The overdot here denotes the derivative with respect to the parameter
$\lambda$. Then the constraints Eq. (\ref{41}) can be rewritten as
\begin{eqnarray}
& &~~~~~~~V_{\mu}V^{\mu} = 0 \nonumber \\
& &\partial_{\mu} \Bigl( e^{- \xi \kappa \phi} V^{\mu}  \Bigr) = 0
\label{45}
\end{eqnarray}
\noindent As we will now show, these three identities suffice to recover the
expression for the classical action. The two constraints have to be treated on
a different footing, though. As can be seen from a simple one dimensional
example, the second constraint, if inserted in the action with a  Lagrange
multiplier, would give rise to a nonlocal theory. Thus, for consistency with
our perturbative approach, as we are disregarding back-reaction, we will only
impose it at the level of equations of motion, and not in the Lagrange
multiplier method.  Therefore, our theory is defined by
\begin{eqnarray}
& & S = \int_{\Gamma} d\lambda \Bigl( V_{\mu} \dot x^{\mu} + \eta V^{\mu}
V_{\mu} \Bigr) \nonumber \\
& &~~~~~~~~~~~~~V_{\mu}V^{\mu} = 0  \\
& & ~~~~~~\partial_{\mu} \Bigl( e^{- \xi \kappa \phi} V^{\mu}  \Bigr) = 0
\nonumber
\label{46}
\end{eqnarray}
\noindent where $\eta$ is the Lagrange multiplier.  The second of these
equations is clearly the $\eta$ equation of motion, as can be seen by direct
variation. The third equation is effectively what introduces the dilaton.
So, varying the action with respect to $V_{\mu}$ gives $V^{\mu} = - \dot
x^{\mu} /2 \eta$, and the differential constraint then becomes
\begin{equation}
\partial_{\mu} \Bigl( e^{- \xi \kappa \phi} \frac{\dot
x^{\mu}}{\eta}\Bigr)_{\Gamma} = 0
\label{47}
\end{equation}
\noindent from which, after noticing that $\partial_{\mu} \dot x^{\nu} = 0$, we
deduce, with help of $\dot x^{\mu} \partial_{\mu} = \frac{d}{d\lambda}$,
\begin{equation}
\frac{d}{d\lambda} \Bigl(\frac{e^{- \xi \kappa \phi}}{\eta} \Bigr)_{\Gamma}= 0
\label{48}
\end{equation}
\noindent or in other words, $\exp(- \xi \kappa \phi) = C \eta$ along the
trajectory $\Gamma$ where $C$ is a (multiplicative) integration constant.
Choosing it to be $C = - 1/4$ we can rewrite the resulting action as
\begin{eqnarray}
& & S = \int_{\Gamma} d\lambda ~e^{\xi \kappa \phi} ~\dot x_{\mu} \dot x^{\mu}
\nonumber \\
& &~~~~~~~~\dot x_{\mu}\dot x^{\mu} = 0
\label{49}
\end{eqnarray}
\noindent where the second equation is the standard dispersion constraint for
massless particles, stating that they move along null lines of the space-time.
The standard variational procedure combined with the null condition then
yields, after simple algebra,
\begin{equation}
\frac{d^2 x^{\mu}}{d\lambda^2} = - \xi ~\kappa ~\partial_{\nu} \phi
{}~\frac{dx^{\nu}}{d\lambda} \frac{dx^{\mu}}{d\lambda}
\label{50}
\end{equation}
\noindent At this point we invoke the Principle of Equivalence to extend this
equation to a curved geometry. We assume that it is to be applied in the
Einstein conformal reference frame, as would be the most natural way for its
extension. Then, the equation of motion becomes
\begin{equation}
\frac{d^2 x^{\mu}}{d\lambda^2} + \Gamma^{\mu}_{\nu\lambda}
\frac{dx^{\nu}}{d\lambda} \frac{dx^{\lambda}}{d\lambda} = - \xi ~\kappa
{}~\partial_{\nu} \phi ~\frac{dx^{\nu}}{d\lambda} \frac{dx^{\mu}}{d\lambda}
\label{51}
\end{equation}

It is illustrative to compare this equation with the ans\"atz of \cite{STW}. We
recall that the field theory defined by Eqs. (\ref{36}-\ref{37}) was derived
from the Maxwell-like
Lagrangian for the $p$-form field strength:
\begin{equation}
S_M = - \frac{1}{p!}\int d^4x \sqrt{g} e^{- \xi \kappa \phi}
{}~F^{\mu\nu\lambda\ldots} F_{\mu\nu\lambda\ldots}
\label{52}
\end{equation}
\noindent Inspection of this Lagrangian shows that a metric field redefinition
(conformal transformation)
\begin{equation}
\bar g_{\mu\nu} = e^{- \alpha ~\xi \kappa \phi} g_{\mu\nu}
\label{53}
\end{equation}
\noindent with $\alpha = 1 /(2 - p)$ ($\alpha = 2 /(D - 2p)$ in $D$ dimensions)
removes any explicit reference to the dilaton in the new frame and renders the
Lagrangian canonical,
\begin{equation}
S_M = - \frac{1}{p!}\int d^4x \sqrt{\bar g} ~\bar F^{\mu\nu\lambda\ldots}
\bar F_{\mu\nu\lambda\ldots}
\label{54}
\end{equation}
\noindent where the overbar refers to the use of the conformally rescaled
metric as the only field to change. Of course, the above is true provided $p
\ne 2$ ($p \ne D/2$ in $D$ dimensions), since the Maxwell Lagrangian density
for this specific ratio is a conformal invariant. If we applied this same
conformal transformation to the connexion of Eq. (\ref{51})
we would obtain after
substitution of the new metric
\begin{equation}
\frac{d^2 x^{\mu}}{d\lambda^2} + \bar \Gamma^{\mu}_{\nu\lambda}
\frac{dx^{\nu}}{d\lambda} \frac{dx^{\lambda}}{d\lambda} = 0
\label{55}
\end{equation}
\noindent in perfect agreement with the ans\"atz of \cite{STW}
for $p \ne 2$ ($p \ne D/2$)! However, our argument would equally well apply to
the $p = 2$ ($p = D/2$)-form field strength, as it is derived without reference
to conformal properties of the associated Lagrangian. Therefore, as a
consequence, the photon's motion is also along the trajectories described by
Eq. (\ref{51}) with the appropriate conformal coupling $\xi$.

Yet the dilaton force appearing explicitly in Eq. (\ref{51})
does not produce any
deviations in the geometry of motion as described in the absence of the
dilaton. It is, indeed, the aforementioned property of null geodesics that they
do not change under conformal transformations which preserves the geometry.
Namely, if we locally rescale the parameter $\lambda$ according to
\begin{equation}
\frac{d\bar \lambda}{d\lambda} = e^{\Omega(\lambda)}
\label{56}
\end{equation}
\noindent with
\begin{equation}
\Omega(\lambda) = - \xi~\kappa \phi(x) \mid_{\Gamma}
\label{57}
\end{equation}
\noindent and note that the null condition is invariant under this
redefinition,
\begin{equation}
g_{\mu\nu}\frac{dx^{\mu}}{d\lambda}\frac{dx^{\nu}}{d\lambda} =
e^{2\Omega} g_{\mu\nu}\frac{dx^{\mu}}
{d\bar \lambda}\frac{dx^{\nu}}{d\bar\lambda} = 0
\label{58}
\end{equation}
\noindent we can rewrite Eq. (\ref{51}) as the null geodesic equation in terms
of the new parameter:
\begin{eqnarray}
\frac{d^2 x^{\mu}}{d\bar \lambda^2}&+& \Gamma^{\mu}_{\nu\lambda}
\frac{dx^{\nu}}{d\bar \lambda} \frac{dx^{\lambda}}{d\bar \lambda} = 0
\nonumber \\
&g_{\mu\nu}&\frac{dx^{\mu}}{d\bar \lambda}\frac{dx^{\nu}}{d\bar\lambda} = 0
\label{59}
\end{eqnarray}
\noindent Here all explicit reference to the dilaton has been absorbed away,
and hence we can conclude that the dilaton force does not affect the geometry
of motion locally. We recall, however, that when the dilaton develops singular
behavior, such as near the black hole singularity,
the transformation (\ref{57})
may be ill defined there. Thence, it can happen that different observers
perceive the singularity differently, according to their proper measure of
length. This point has been well illustrated by \cite{STW} in their discussion
of proper motion of axions and dilatons in the dilaton black hole background of
the truncated theory. As a corollary we deduce that the spherically symmetric
background with the dilaton can not be distinguished from the dilaton-less case
on the basis of geometric aberrations, and hence both backgrounds would appear
the same in experiments such as the standard tests of General Relativity.

It should be obvious, though, that the dilaton force does produce observable
physical effects in a locally inertial frame, although the geometry is
independent of it. We offer the following simple illustration of this claim.
Consider the Schwarzschild background to $O(\alpha')$ with the dilaton and
radial motion in it. The action describing  dynamics of the probe is, in the
Einstein frame,
\begin{equation}
S = \int_{\Gamma} d\lambda ~e^{\xi \kappa \phi} ~\Bigl(
\frac{1}{1 - (2GM/r)} \bigl(\frac{dr}{d\lambda})^2 -
\bigl(1 - \frac{2GM}{r}\bigr) \bigl(\frac{dt}{d\lambda}\bigr)^2 \Bigr)
\label{60}
\end{equation}
\noindent supplemented with the null condition. The four-velocity of the probe
is
\begin{equation}
\frac{dx^{\mu}}{d\lambda} = u^{\mu} = \left(\matrix{\frac{E}{1 - (2GM/r)}
e^{-\xi \kappa \phi}\cr
E e^{-\xi \kappa \phi}\cr
{}~0\cr
{}~0\cr}\right)
\label{61}
\end{equation}
\noindent where $E$ is the energy integral of the probe. The stationary
observer at the distance $r$ from the origin, whose space-time evolution is
described by the four-velocity
\begin{equation}
v^{\mu} = \left(\matrix{\frac{1}{\sqrt{1 - (2GM/r)}} \cr
{}~0\cr
{}~0\cr
{}~0\cr}\right)
\label{62}
\end{equation}
\noindent detects the energy of the probe in the gravitational-dilatonic well
as
\begin{equation}
h\nu = v_{\mu} u^{\mu} = \frac{E}{\sqrt{1 - (2GM/r)}} e^{-\xi \kappa \phi}
\label{63}
\end{equation}
\noindent Hence, the redshift as seen by an observer stationary in the
Einstein frame will clearly be affected by the dilaton. Therefore, the dilaton
indeed does produce observable effects. In passing, we note that the redshift
formula,  as dictated by the above,
\begin{equation}
\frac{\nu_1}{\nu_2} = Z_{GR}~e^{\xi \kappa \bigl(\phi(2) - \phi(1)\bigr)}
\label{64}
\end{equation}
\noindent where $Z_{GR}$ is the redshift in standard GR, is not invariant under
conformal transformation. This is a consequence of the fact that observers
normally move along time-like geodesics, which are affected by confomal
transformations. Thus, the formula above is clearly observer-dependent. This
results in the need for a careful definition of local physics (energy-momentum
relationships) in presence of the dilaton hair. For example, although the
geometric picture obtained with our approach coincides with that of \cite{STW},
the formula for the redshift which one might attempt to obtain using the
expressions for the conformal transform of the metric where the dilaton force
is absent would differ from ours, since it would \underbar{not} correspond to
the Einstein stationary observer, but to its conformal transform. Numerically,
the correction to the GR redshift for light emitted by the Sun is
$1 + O(1/G M_\odot R_\odot)$ and so amounts to a correction of one part in
$10^{79}$.

In a more general situation, when the black hole is charged, the dilaton force
would not just affect the dispersion relations, but would also affect the
Lorentz force acting on charged particles. However, we will not discuss this
case in detail, as the effects in question would be incredibly small, in the
absence of a highly charged (i.e. almost critical) source as may be seen from
our numerical results of the previous section.

In the remainder of this section we undertake a brief discussion of the other
class of induced forces specific to the superstring models of gravity. These
are the effects induced by the presence of the Kalb-Ramond axion. One such
effect, the rotation of the photon polarization vector in a background with
axion hair, has been pointed out previously \cite{CDKO2,DKO}. Again we will
simplify our gedanken experiment to the maximum possible, so we can discuss
effects of the axion hair alone, and not as coupled to the dilaton hair. Hence
we will
assume absence of the dilaton from the theory altogether, presuming that it had
decoupled dynamically by a symmetry breaking type mechanism which provided it
with a large mass. Such a theory has been noted to posses secondary axion hair
on black holes, and more generally, on all rotating, or dyonic matter sources.
Therefore, it could be possible, in principle, to conduct experiments testing
hair without explicit reference to black holes. For the sake of simplicity, we
will take this possibility here. We will assume that the axion hair arises from
a nonvanishing dyon anomaly source in a flat Minkowski background. In this way
we will avoid the unnecessary complication of having to worry about the
non-trivial metric in our equations.
We expect that our results will be qualitatively unchanged in a curved
background, in view of the fact that we assume the validity of the Principle of
Equivalence in the Einstein conformal frame.

Our intention is to study the influence of the axion hair on the motion of
photons. For this purpose, we can concentrate solely on the (modified)
Maxwell's equations describing the photon, as we disregard the backreaction
assuming that the contribution of the photon to the action is miniscule
compared to the background fields. The linearity of Maxwell's equations
with the anomalous axion coupling is consistent with this treatment, as it
splits the photon from the background electromagnetic field. We note that the
background field and the axion hair are given by the dyon solution of Ref.
\cite{CKO1,LW}. Then, we will assume that the observations are being made far
away from the
dyon. Far away from the dyon, the variation of the gradient of the axion field
is negligible compared to the short distance physics of the experiment.
Therefore, we can replace $\partial_{\lambda}a = q_{\lambda} = (0, 0, 0, a')$
where $q_{\lambda} \approx const$ and we have oriented the local coordinate
system so that the gradient is along the vertical axis. The independent
equations of motion for the photon are
\begin{equation}
\partial_{\mu}  F^{\mu\nu} - \frac{\sqrt{2}}{2} \kappa
\epsilon^{\nu\lambda\sigma\rho} q_{\lambda} F_{\sigma\rho} = 0
\label{65}
\end{equation}
\noindent with $F_{\mu\nu} = \partial_{\mu} A_{\nu} - \partial_{\nu} A_{\mu}$,
and the gauge condition $\partial_{\mu} A^{\mu} = 0$. Note that here we employ
the metric $\eta = (1, -\vec 1)$ of signature $-2$ more customary to particle
physics, to render the results more transparent. Normally, in the absence of
the anomalous coupling, the residual gauge invariance under gauge
transformations which solve the free Klein-Gordon equation, allows us to set
$A_0 = 0$ and hence employ the Coulomb gauge. However this is not possible
here, as due to the anomaly the Coulomb gauge is not unitary and the associated
ghosts would not decouple. This can be seen as follows. If we gauge transform
$A^{\mu}$ to $A'^{\mu} = A^{\mu} + \partial^{\mu} \tilde \Lambda$ such that
$A'^{0} = 0$, the gauge transformation $\tilde \Lambda$ would have to satisfy
the two equations
\begin{eqnarray}
&& \dot {\tilde \Lambda} = -A^{0} \nonumber \\
&& \partial^2 \tilde \Lambda = 0
\label{66}
\end{eqnarray}
\noindent The first equation is solved by
\begin{equation}
\tilde \Lambda(t, \vec x) = - \int^{t}_{t_0} d\tau A^0(\tau, \vec x)
+ f(\vec x)
\label{67}
\end{equation}
\noindent where $f(\vec x)$ is some function independent of time to be
determined by the boundary conditions. Application of the D'Alembertian to the
solution yields
\begin{equation}
\partial^2 \tilde \Lambda(t, \vec x) = - \dot A^0(t, \vec x) + \int^{t}_{t_0}
d\tau {\vec \nabla}^2 A^0(\tau,\vec x) + {\vec \nabla}^2 f(\vec x)
\label{68}
\end{equation}
\noindent where $\vec \nabla^2$ is the spatial Laplacian. Using the equations
of motion for $A^0$, after some tedious algebra Eq. (\ref{68}) can be rewritten
as
\begin{equation}
\partial^2 \tilde \Lambda(t, \vec x) = - \dot A^0(t_0, x) + {\vec \nabla}^2
f(\vec x) + \sqrt{2} \kappa \vec q \cdot \int^{t}_{t_0} d\tau \vec \nabla
\times \vec A(\tau, \vec x)
\label{69}
\end{equation}
\noindent Recalling that $\vec A' = \vec A - \vec \nabla \tilde \Lambda$, we
note that $\vec A$ in the integrand above can be substituted with $\vec A'$.
Defining $\chi = \partial^2 \tilde \Lambda$ we see that
\begin{equation}
\dot \chi =  \sqrt{2} \kappa \vec q \cdot \vec \nabla \times \vec A'
\label{70}
\end{equation}
\noindent which in general does not vanish. Hence $\partial^2 \tilde \Lambda$
is generally not constant in time and $\tilde \Lambda$ will not represent a
legitimate gauge transformation. In other words, there will have to be
additional terms included in the equations of motion accounting for the
variation of $\tilde \Lambda$, which from the analogy with the BRST method will
correspond exactly to the gauge field - ghost couplings. In the special case
when $\vec q \cdot \vec \nabla \times \vec A' = 0$ , the Coulomb gauge will be
admissible. Then $\vec q \perp \vec \nabla \times \vec A'$, since $\vec \nabla
\times \vec A' = 0$ together with the Coulomb gauge condition $\vec \nabla \vec
A' = 0$ uniquely sets $\partial_{k}\vec A' = 0$. But this corresponds to a
homogeneous background electric field which can be removed by boundary
conditions at infinity. We note in passing that in the treatment of \cite{CFJ}
it was assumed that the vector $q_{\mu}$ is timelike. Thus in its rest frame
$\vec q = 0$ and the last term in Eq. (\ref{69})
is absent. Therefore the Coulomb
gauge is unitary and does belong to the Lorenz class. This is the reason why
those authors were able to use it. We remark, however, that at least due to the
axion hair it is more natural to expect that the vector $q_{\mu}$ will be
spacelike, as above.

We look for the eigensolutions of the Eq. (\ref{65}), i.e. the possible
\underbar{physical} polarization states of the photon in such background. The
residual gauge invariance in combination with the fact that we are already
working with field configurations belonging to the Lorenz class tells us that
there should be only two independent polarization states for each direction of
motion. However, as we noted in the previous paragraph, they are not transverse
in general. They could be found in two different ways. Either one could find
another gauge condition which intersects non-trivially the Lorenz class and
propagates freely and then find all polarization states satisfying those two
constraints, or one could find the most general solution from the Lorenz class
and then gauge away the pure gauge contributions with help of the residual
gauge invariance. We adopt the second approach, or in other words, we choose to
work with the Fermi Lagrangian:
\begin{equation}
S = \int d^4x \Bigl( - \frac{1}{4} F_{\mu\nu}F^{\mu\nu} - \frac{1}{2\zeta}
\bigl(\partial_{\mu} A^{\mu} \bigr)^2 + \frac{\sqrt{2}\kappa}{4}
\epsilon^{\mu\nu\lambda\sigma} q_{\mu}A_{\nu}F_{\lambda\sigma} \Bigr)
\label{71}
\end{equation}
\noindent The Lagrangian which reproduces the Eq. (\ref{65})
with the Lorenz gauge
condition is given by $\zeta = 1$. The equations of motion are
\begin{eqnarray}
&& \partial^2 A^{\mu} - (1 - \frac{1}{\zeta}) \partial^{\mu} \partial \cdot A -
\sqrt{2} \kappa \epsilon^{\mu\nu\lambda\sigma}
q_{\nu}\partial_{\lambda}A_{\sigma} = 0 \nonumber \\
&& ~~~~~\partial^2 \partial_{\mu} A^{\mu} = 0
\label{72}
\end{eqnarray}
\noindent Hence the Lorenz gauge constraint is a free field and its value is
completely determined by the boundary conditions. The evolution operator of the
electromagnetic field is
\begin{equation}
K^{\mu\nu} = \eta^{\mu\nu}\partial^2  - (1 - \frac{1}{\zeta}) \partial^{\mu}
\partial^{\nu} - \sqrt{2} \kappa \epsilon^{\mu\nu\lambda\sigma}
q_{\lambda}\partial_{\sigma}
\label{73}
\end{equation}
\noindent and therefore the propagator in the momentum picture is
\begin{eqnarray}
G_{\mu\nu} &=& - \frac{{\rm i}}{p^2 + 2\kappa^2 q^2 - 2\kappa^2 \frac{(q \cdot
p)^2}{p^2} + {\rm i} \epsilon}\Bigl(\eta_{\mu\nu} + \bigl(2\kappa^2
\frac{(q\cdot p)^2}{(p^2 + {\rm i}\epsilon)^2} - 1
\bigr)\frac{p_{\mu}p_{\nu}}{p^2 + {\rm i}\epsilon} \nonumber \\
& & + 2\kappa^2 \frac{q_{\mu}q_{\nu}}{p^2 + {\rm i}\epsilon} - 2\kappa^2
\frac{q\cdot p}{(p^2 + {\rm i}\epsilon)^2}\bigl( q_{\mu}p_{\nu} +
p_{\mu}q_{\nu}\bigr)
 + \sqrt{2}{\rm i}\kappa \epsilon_{\mu\nu\lambda\sigma} \frac{q^{\lambda}
p^{\sigma}}{p^2 + {\rm i}\epsilon} \Bigr) - {\rm i} \zeta
\frac{p_{\mu}p_{\nu}}{p^2 + {\rm i}\epsilon}
 \label{74}
\end{eqnarray}
\noindent Since for a spacelike vector $q_{\mu}$, $m^2 = - 2\kappa^2 q^2 \ge
0$, the physical poles of the propagator are determined by the roots of $p^4 -
m^2p^2 - 2\kappa(q \cdot p)^2 = 0$. This equation states that in addition to
massless modes there will appear massive modes in the photon spectrum. The
parameter $m^2$ will measure the photon mass. In general, the mass will be
polarization-dependent, and will have to be obtained from diagonalization of
an operator matrix much like an of inertia tensor. We will illustrate this in
the special case of a photon moving in the tangent plane, orthogonally to $\vec
q$.

In the Feynman gauge $\zeta = 1$ and with boundary conditions s.t. $\partial
\cdot A = 0$, we can rewrite the Fourier modes of Eq. (\ref{65}) as
\begin{eqnarray}
\ddot \phi &+& \vec p^{~2}\phi - \sqrt{2}{\rm i} \kappa \vec q \cdot (\vec p
\times \vec A) = 0 \nonumber \\
\ddot {\vec A} &+& \vec p^{~2} \vec A - \sqrt{2} \kappa \vec q \times \dot{\vec
A} - \sqrt{2}{\rm i} \kappa \vec q \times \vec p ~\phi = 0
\label{75}
\end{eqnarray}
\noindent where overdot denotes derivative with respect to time, and the
solution is of the form
$A^{\mu} = (\phi(t), \vec A(t)) \exp({\rm i} \vec p \cdot \vec x)$. The Lorenz
gauge condition reads $\dot \phi + {\rm i} \vec p \cdot \vec A = 0$. With its
help we obtain, after taking its time derivative, and using the $\phi$ equation
of motion,
\begin{equation}
\phi = \frac{\rm i}{\vec p^{~2}} \Bigl( \sqrt{2} \kappa  \vec q \cdot (\vec p
\times \vec A) + \vec p \cdot \dot{\vec A }\Bigr)
\label{76}
\end{equation}
\noindent and
\begin{equation}
\ddot {\vec A} - \sqrt{2}\kappa \Bigl( \vec q \times \dot{\vec A} - \frac{\vec
q \times \vec p}{\vec p^{~2}}(\vec p \cdot \dot{\vec A })\Bigr)
+ \vec p^{~2} \vec A + 2\kappa^2 \frac{\vec q \times \vec p}{\vec
p^{~2}}\Bigl(\vec q \cdot (\vec p \times \vec A)\Bigr) = 0
\label{77}
\end{equation}
\noindent The equation of motion for $\vec A$ can be rewritten in the matrix
form
\begin{equation}
\ddot A^j + M^{jk}\dot A^k + N^{jk}A^k = 0
\label{78}
\end{equation}
\noindent where in the coordinate frame $q_{\mu} = (0, 0, 0, q)$ the matrices
are
\begin{equation}
M = - \sqrt{2}\kappa q \left(\matrix{
\frac{p_1p_2}{\vec p^{~2}}&\frac{p_2^2}{\vec p^{~2}}-1&\frac{p_2p_3}{\vec
p^{~2}}\cr
1-\frac{p_1^2}{\vec p^{~2}}&-\frac{p_1p_2}{\vec p^{~2}}&-\frac{p_1p_3}{\vec
p^{~2}}\cr
{}~0~&~0~&~0~\cr}\right)
\label{79}
\end{equation}
\noindent and
\begin{equation}
N = \left(\matrix{
\vec p^{~2} + 2\kappa^2 q^2 \frac{p_2^2}{\vec p^{~2}}&- 2\kappa^2
q^2\frac{p_1p_2}{\vec p^{~2}}&~0~\cr
- 2\kappa^2 q^2\frac{p_1p_2}{\vec p^{~2}}&\vec p^{~2} + 2\kappa^2
q^2\frac{p_1^2}{\vec p^{~2}}&~0~\cr
{}~0~&~0~&~0~\cr}\right)
\label{80}
\end{equation}

Solving the system is a standard exercise in simultaneous diagonalization of
two quadratic forms. The situation here is similar to the case of small
oscillations with dissipation. This problem in general is not tractable;
however, here the matrix coefficient of the highest derivative is unity, and
hence the system is diagonalizable. Thus, the standard prescription $A^j =
A_0^j
\exp(-{\rm i} \omega t)$ yields the secular equation
\begin{equation}
\det C(\omega) = 0
\label{81}
\end{equation}
\noindent where $C(\omega) = N - {\rm i} \omega M - \omega^2$. As mentioned
above, $C$ is very much like an inertia tensor, and its null vectors define the
independent polarizations for a given direction of motion. For brevity we will
not discuss the most general solution of the system above. After all, the
system does represent only an approximate dynamical model and for the actual
arbitrary motion of the photon the variation of the axion will come into play.
Yet, if one conducts an experiment in the lab, the results will indicate an
effective mass term for the photon. To illustrate this, consider a photon
moving in a locally horizontal direction and orient the coordinate axes
so that the $x$ axis points in the direction of motion. Then $p_2 = p_3 = 0$.
The evolution matrix $C(\omega)$ becomes
\begin{equation}
C(\omega) = \left(\matrix{
\vec p^{~2} - \omega^2& - {\rm i}\sqrt{2}\kappa q\omega&~0~\cr
{}~0~&\vec p^{~2} - \omega^2 + 2\kappa^2 q^2 \frac{p_1^2}{\vec p^{~2}}&~0~\cr
{}~0~&~0~&\vec p^{~2} - \omega^2 \cr}\right)
\label{82}
\end{equation}
\noindent and so the secular equation is, recalling $m^2 = - 2\kappa q_{\mu}
q^{\mu} = 2\kappa q^{2}$,
\begin{equation}
\Bigl(\vec p^{~2} - \omega^2\Bigr)^2\Bigl(\vec p^{~2} - \omega^2  + m^2 \Bigr)
= 0
\label{83}
\end{equation}
\noindent The eigenfrequencies are $\omega^2 = \vec p^{~2}$ (double root) and
$\omega^2 = \vec p^{~2} + m^2$ (single root). The last formula illustrates the
identification of $m^2$ as the photon mass. The associated polarization vectors
can be computed as the null vectors of $C(\omega)$ for the corresponding
eigenfrequencies. We will skip the laborious (but straightforward) algebra and
just quote the final solution:
\begin{eqnarray}
\phi &=& U_1 e^{{\rm i}p(x - t)} - U_2 e^{{\rm i}p(x + t)}  - {\rm i}
\frac{p}{m} V_1 e^{{\rm i}(px - \omega_m t)} + {\rm i}\frac{p}{m}
V_2 e^{{\rm i}(px + \omega_m t)} \nonumber \\
\vec A &=& \Bigl(U_1 e^{{\rm i}p(x - t)} + U_2 e^{{\rm i}p(x + t)}\Bigr)
\vec e^{~1} + \Bigl(W_1 e^{{\rm i}p(x - t)} + W_2 e^{{\rm i}p(x + t)}\Bigr)
\vec e^{~3} \\
&+& \Bigl( V_1 e^{{\rm i}(px - \omega_m t)} - V_2 e^{{\rm i}(px + \omega_m
t)}\Bigr) \vec e^{~2} - {\rm i} \frac{\omega_m}{m} \Bigl( V_1 e^{{\rm i}(px -
\omega_m t)} + V_2 e^{{\rm i}(px + \omega_m t)}\Bigr)\vec e^{~1}
\nonumber
\label{84}
\end{eqnarray}
\noindent where $\omega_m = \sqrt{p^2 +
m^2}$. The $U_{1,2}$ terms are pure gauge. To see that, one can apply the gauge
transformation $\Lambda = - \frac{{\rm i}}{p}\Bigl(U_1 e^{{\rm i}p(x - t)} +
U_2 e^{{\rm i}p(x + t)}\Bigr)$ which propagates freely:
$\partial^2 \Lambda = 0$, to remove these degrees of freedom. Hence, the
general physical solution is
\begin{eqnarray}
\phi &=&  - {\rm i} \frac{p}{m} V_1 e^{{\rm i}(px - \omega_m t)}
+ {\rm i}\frac{p}{m} V_2 e^{{\rm i}(px + \omega_m t)} \nonumber \\
\vec A &=& \Bigl( V_1 e^{{\rm i}(px - \omega_m t)} - V_2 e^{{\rm i}(px +
\omega_m t)}\Bigr) \vec e^{~2} - {\rm i} \frac{\omega_m}{m} \Bigl( V_1 e^{{\rm
i}(px - \omega_m t)} + V_2 e^{{\rm i}(px + \omega_m t)}\Bigr)\vec e^{~1} \\
&+& \Bigl(W_1 e^{{\rm i}p(x - t)} + W_2 e^{{\rm i}p(x + t)}\Bigr) \vec e^{~3}
\nonumber
\label{85}
\end{eqnarray}
\noindent It is interesting to note that there is a longitudinal
component of the vector potential, as carried by the mass. However, the
solution still contains only two independent polarizations, which is a
manifestation of the (unbroken) gauge invariance of the theory even in the
presence of mass. Note that in the limit $m \rightarrow 0$ the longitudinal
massive component transmutes into a pure gauge and can be gauged away. The
resulting expression would be just the standard formula from ordinary classical
EM in Coulomb gauge.

The explicit formula for the mass can be easily obtained by substituting $\vec
q^{~2} = (\vec \nabla a )^2$ for appropriate axion field hair. In our case, the
full $O(\alpha')$ axion hair is determined by formula (A13). To get the
photon mass, we could truncate it and retain only the dyon contribution,
containing the cross term between the electric and magnetic charges. However,
the order-of-magnitude estimate will be served just as well by the contribution
due to the LCS source due to the rotation, for an experiment on Earth's
surface. Here we present formulas for both cases, and calculate the mass
numerically as seen by experiments on Earth, for comparison. The masses are,
after we recover the units, and use $A = 2 MR^2 \omega /5$ for the angular
momentum of Earth, where $\omega_E$ the angular velocity of rotation,
\begin{eqnarray}
& & m^2_1 = \frac{4\pi^2\hbar^4}{c^4 g^4} \frac{N^2}{M_E^2 R_E^4} \nonumber \\
& & m^2_2 = \frac{16\pi^2 G \hbar^3}{c^7 g^4}\frac{\omega^2_E}{R_E^2}
\label{86}
\end{eqnarray}
\noindent where they correspond to the dionic and rotating sources,
respectively. The string tension $\alpha'$ is given by $\alpha' = 2
\kappa^2/g^2$ where $g$ is the unification scale coupling constant with the
typical magnitude $g^2 \sim 0.1$. We have used the Dirac quantization condition
$Q_eQ_m = \hbar c N/2$ for the dyon, and computed the dipole field mass near
the equator, where it would be the largest. Here $N$ would measure the total
charge of Earth in the units of electron charge. The two numerical values for
the masses on the surface of Earth are $m_1 \sim 2 \times 10^{-92} {\rm MeV}$,
$m_2 \sim 4 \times 10^{-65} {\rm MeV}$, well within the limit $m_{\gamma} < 3
\times 10^{-33} {\rm MeV}$. The rotation-generated mass is significantly bigger
than the dyonic one, and what's even more interesting, does not depend on the
mass of the celestial body which supports the axion hair. Actually, for a body
of the radial size $10^3$ times smaller than Earth, which rotates much faster
than it ($\omega /\omega_E \sim 10^{8}$), the photon mass generated by the
axion hair approaches the experimental limit, but remains well within it as it
still lags by $21$ orders of magnitude. Thus, even the experiments involving
compact rapidly rotating objects such as milisecond pulsars would not disclose
the
mass of the photon above the experimental limit set by the measurements on
Earth.

We conclude this discussion by noting the origin of the constraints on the
mass. They can be divided in two classes: the geomagnetic and the cosmological.
Both were discussed in \cite{CFJ} and would remain much the same here. For our
purposes, however, the geomagnetic constraints are more sensible, as our
approximation breaks down when experiments involving photons are conducted over
large distances, since the variation of the axion field can not be neglected.
In closing, we would like to stress that the existence of a very small photon
mass could be viewed as an argument in favor of existence of axions, or an
axion-like phenomenon.


\vskip.3in
{\bf Conclusion}
\vskip.3in

In this paper we have analyzed physical properties of some of the recently
constructed 4$D$ black hole solutions in string theory. We have discussed
similarities and differences between some of the exact solutions of the
classical effective action, describing dynamics of the model-independent
degrees of freedom which appear in the supergravity and superstring theories.
Our focus was on the importance of the higher order corrections in the string
tension $\alpha'$ and higher genus contributions, and we find that in general
these terms may render significant departures from the observed classical
behavior. We have also demonstrated that in string perturbation theory the
solutions are essentially equivalent with the approximate $O(\alpha')$
solutions we have constructed previously.

Finally, we have discussed observable consequences of the black hole hair. We
have found that the hair indeed yields departures from Einstein's General
Relativity. In general, there would appear forces derived from the hair which
may influence the geometry  of motion as perceived by point-like test particles
in such background, as well as induce different energy-momentum relations from
standard GR. In particular, we have demonstrated that the dilaton hair results
in different red-shift relations for probes moving in gravity-dilaton potential
wells. Although for the specific case considered the dilaton force does not
produce any deviations in the geometry of motion of massless particles (thus
being indistinguishable from standard GR by radar echo delay and deflection of
light experiments) we have argued for the occurence of such deviations in the
case of massive and/or charged probes. In addition, we have addressed the issue
of nontrivial axionic hair and have found that in the regime of small fields,
its anomalous coupling to the electro-magnetic field could provide the photon
with an effective topological mass term.

\vskip.3in
\noindent {{\bf Acknowledgements}}
\vskip.3in

\noindent We would like to acknowledge useful conversations with
Don Page. The work of BAC and NK
was supported in part by the Natural Sciences
and Engineering Research council of Canada.
The work of KAO was supported in
part by DOE grant DE-AC02-83ER-40105, and by a
Presidential Young Investigator
Award. BAC and KAO would like to thank the CERN Theory Division for kind
hospitality during part of this research.

\vfill
\eject
\newpage

\appendix
\renewcommand{\theequation}{A\arabic{equation}}
\setcounter{equation}{0}
{\bf Appendix}
\vskip.3in

Here we collect the various metric, dilaton and axion solutions. We also
indicate all of the appropriate field and charge redefinitions as necessary.

The action in \cite{Gibb,GM,GHS} is
\begin{equation}
S = \int d^4x \sqrt{g} \Bigl\{\frac{R}{2 \kappa^2} - \frac{1}{\kappa^2}
\partial_{\mu}\phi' \partial^{\mu}\phi' - \frac{1}{2\kappa^2} e^{-2\phi'} F'^2
\Bigr\}
\label{A1}
\end{equation}
\noindent (we have multiplied the action in
\cite{Gibb,GM,GHS} by $-1 /2 \kappa$). It dictates the dilaton field
redefinition
\begin{equation}
\phi' = \frac{ \kappa \phi}{\sqrt{2}}
\label{A2}
\end{equation}
\noindent and the charge in which $F'^2$ is expressed in ($F'^2 \propto Q'^2$)
is
\begin{equation}
Q'^2 = \alpha' \frac{Q^2}{8}
\label{A3}
\end{equation}
\noindent Similarly, the Sen action is written as
\begin{equation}
S = \int d^4x \sqrt{g} \Bigl\{\frac{R}{2 \kappa^2} - \frac{1}{24\kappa^2} e^{-2
\tilde \phi} \tilde H^2 - \frac{1}{4\kappa^2} \partial_{\mu}\tilde \phi
\partial^{\mu}\tilde \phi - \frac{1}{16 \kappa^2} e^{-\tilde \phi}\tilde F^2
\Bigr\}
\label{A4}
\end{equation}
\noindent (we have multiplied the Sen action by $1 / 2\kappa^2$ and performed
the Weyl rescaling), so that
\begin{equation}
\tilde \phi = 2 \phi' = \sqrt{2}\kappa \phi
\label{A5}
\end{equation}
\begin{equation}
\tilde H_{\mu\nu\lambda} = 12 \kappa H_{\mu\nu\lambda}
\label{A6}
\end{equation}
\begin{equation}
\tilde Q^2 = Q'^2 = \alpha' \frac{Q^2}{8}
\label{A7}
\end{equation}
\noindent (note that $\tilde F \propto 2 \sqrt{2} \tilde Q$ whereas $F,F'
\propto Q,Q'$).

The Kerr-Newmann solution can be written as
\begin{eqnarray}
ds^2 &=& \Bigl(
{r^2 + A^2 \cos^2 \theta - 2GMr + Gq^2 \over r^2 + A^2 \cos^2 \theta}
\Bigr)~dt^2 - \Bigl({r^2 + A^2 \cos^2 \theta  \over r^2 + A^2 - 2GMr +
Gq^2}\Bigr)~dr^2  \nonumber \\
&-& \Bigl( r^2 + A^2 \cos^2 \theta \Bigr)~d\theta^2 - 2 A^2 \sin^2 \theta
\Bigl({ 2GMr - Gq^2 \over r^2 + A^2 \cos^2 \theta}
\Bigr)~dtd\phi \\
&-& \Bigl(
{(r^2 + A^2)^2 - A^2 \sin^2 \theta (r^2 + A^2 - 2GMr + Gq^2 \over r^2 + A^2
\cos^2 \theta} \Bigr)~\sin^2\theta d\phi^2  \nonumber
\label{A8}
\end{eqnarray}
\noindent and
\begin{equation}
A_{\mu} = \Bigl(\bigl( {Q_e r \over r^2 + A^2 \cos^2 \theta}\bigr),~0,~0,
{}~-\bigl( {Q_e r A\sin^2 \theta \over r^2 + A^2 \cos^2 \theta}\bigr) \Bigr)
\label{A9}
\end{equation}
\noindent where the charge in the metric is given by $q^2 = (\pi \alpha' /
\kappa^2) Q^2$ in terms of the charge given in $A_{\mu}$ and hence in
$F_{\mu\nu}$. The metric given in \cite{Gibb,GM,GHS} is (they did not consider
the rotating case)
\begin{equation}
ds^2 = \bigl(1 - \frac{2GM}{\rho}\bigr) dt^2 - \bigl(1 -
\frac{2GM}{\rho}\bigr)^{-1} d\rho^2 - \rho \bigl(\rho - \frac{Q'^2}{GM}\bigr)
d\Omega
\label{A10}
\end{equation}
\noindent and that of Sen
\begin{eqnarray}
ds^2 &=& -\frac{\tilde \rho^2 + A^2\cos^2\theta -2m\tilde \rho}{ \tilde \rho^2
+A^2\cos^2\theta +2m\tilde \rho\sinh^2\frac{\alpha}{2}} dt^2
+ \frac{\tilde \rho^2
+A^2\cos^2\theta +2m\tilde \rho\sinh^2\frac{\alpha}{2}}{ \tilde \rho^2 +A^2
-2m\tilde \rho}
d\tilde \rho^2 \nonumber ~~~~~~~~~~~~~~~~~~~~~~~~~~~~~\\
& &+(\tilde \rho^2
+A^2\cos^2\theta +2m\tilde \rho\sinh^2\frac{\alpha}{2}) d\theta^2
-\frac{4m\tilde \rho A\cosh^2\frac{\alpha}{2}\sin^2\theta}{\tilde \rho^2
+A^2\cos^2\theta +2m\tilde \rho\sinh^2\frac{\alpha}{2}} dtd\phi\\
& &+\Bigl((\tilde \rho^2+A^2)(\tilde \rho^2+A^2\cos^2\theta) +2m\tilde \rho
A^2\sin^2\theta +4m\tilde \rho
(\tilde \rho^2+A^2) \sinh^2\frac{\alpha}{2}+ 4m^2\tilde
\rho^2\sinh^4\frac{\alpha}{2}\Bigr)\nonumber \\
& &\times \frac{\sin^2\theta}{\tilde \rho^2
+A^2\cos^2\theta +2m\tilde \rho\sinh^2\frac{\alpha}{2}} d\phi^2 \nonumber
\label{A11}
\end{eqnarray}
\noindent where $\sinh^2(\alpha /2) = \tilde Q^2 / (2G^2M^2 - \tilde Q^2)$, $2m
= (2G^2M^2 - \tilde Q^2) / GM $.

We turn now to the dilaton and axion solutions; our $O(\alpha')$ solution is
\cite{CKO2}, for the dilaton,
\begin{eqnarray}
\phi(r,\theta)&=&\frac{\alpha\acute{ }\sqrt{2}}{16\kappa}
([\frac{-1}{GMr}[2(1+\frac{GM}{r}+\frac{4G^2M^2}{3r^2})
+(Q_e^2-Q_m^2)] \nonumber \\
&+& A^2[\frac{1}{2G^3M^3r}(1+\frac{GM}{r}+\frac{8G^2M^2}{3r^2}
+\frac{6G^3M^3}{r^3}+\frac{64G^4M^4}{5r^4})+\frac{Q_e^2-Q_m^2}{3GMr^3}]
+\ldots]P_0(\cos\theta)\nonumber \\
&+&[\frac{2Q_eQ_mA}{GMr^2}+\ldots]P_1(\cos\theta)\\
&+&[\frac{2A^2}{3GMr^3}[\frac{14}{5}(1+\frac{3GM}{r}+\frac{48G^2M^2}{7r^2})
+(Q_e^2-Q_m^2)]+\ldots]P_2(\cos\theta)+\ldots) \nonumber
\label{A12}
\end{eqnarray}
\noindent and the axion
\begin{eqnarray}
a(r,\theta)&=&\frac{6\alpha\acute{ }\sqrt{2}}{8\kappa}
([-\frac{Q_eQ_m}{6GMr}(1-\frac{A^2}{3r^2}+\ldots)]P_0(\cos\theta)
\nonumber \\
&+&[\frac{A}{12GMr^2}[-\frac{5}{2}(1+\frac{2GM}{r}+\frac{18G^2M^2}{5r^2})
-(Q_e^2-Q_m^2)]+\ldots]P_1(\cos\theta)\\
&+&[\frac{Q_eQ_mA^2}{9GMr^3}+\ldots]P_2(\cos\theta)+\ldots)
\nonumber
\label{A13}
\end{eqnarray}
\noindent where contributions to dilaton from electric and magnetic charges,
rotation and Gauss-Bonnet curvature terms have been included. The
dilaton solution in \cite{Gibb,GM,GHS}
for a nonrotating electrically charged black hole is
\begin{equation}
\phi' = \frac{1}{2} \log\bigl(1 - \frac{Q'^2}{GM\rho}\bigr)
\label{A14}
\end{equation}
\noindent and for a rotating eletrically charged black hole \cite{Sen} is
\begin{equation}
\tilde \phi = - \log\bigl(\frac{\tilde \rho^2 + \frac{\tilde Q^2}{GM}
\tilde \rho + A^2 \cos^2 \theta}{\tilde \rho^2 + A^2 \cos^2 \theta}\bigr)
\label{A15}
\end{equation}
\noindent When expanded to $O(\alpha')$, with the coordinate transformations
Eq. (\ref{12}) and conventions of Eq. (\ref{A5}) to (\ref{A7}), these two
solutions are identical with the purely electric contribution in Eq. (A12).

Now we contrast the axion solutions to ours. In the case of \cite{GHS} the
axion is zero identically since they do not include the LCS source and work
with purely electric (or magnetic) configurations. Before we can compare the
Sen solution with ours, we need to dualize it first. The axion solution of Sen
is expressed in terms of the Kalb-Ramond $2$-form
\begin{equation}
\tilde B_{t\phi} = {2m\tilde \rho A\sinh^2{\alpha\over 2}\sin^2\theta \over
\tilde \rho^2
+A^2\cos^2\theta +2m\tilde \rho\sinh^2{\alpha\over 2}}
\label{A16}
\end{equation}
\noindent We need to dualize it to the pseudoscalar axion which we wrote the
solutions in terms of. This can easily be done with the following formulas.
Sen's normalizations are such that
\begin{equation}
\tilde H_{\mu\nu\lambda} = \partial_{\lambda}\tilde B_{\mu\nu} - \frac{1}{4}
\tilde A_{\lambda}\tilde F_{\mu\nu} + ({\rm cyclic~permutations})
\label{A17}
\end{equation}
\noindent where the gauge terms must be included in the Chern-Simons
combination since the twisting procedure which Sen employs convolutes the
electric charge with the angular momentum producing a non-vanishing gauge
anomaly. We have noticed this effect previously, when working in a rotating
background to order $O(\alpha')$ \cite{CKO2}. Then from the action, the $\tilde
H_{\mu\nu\lambda}$ is related to our
$H_{\mu\nu\lambda}$  by (\ref{A6}) or in terms of forms,
\begin{equation}
\tilde H = 2\kappa H
\label{A18}
\end{equation}
\noindent Our conventions for the wedge product are defined by $\alpha \wedge
\beta = Alt ~\alpha \otimes \beta$, whereas Sen uses $\alpha \wedge \beta =
\frac{(p + q)!}{p! q!} Alt ~\alpha \otimes \beta$, with $p$ and $q$ the ranks
of the two forms. In the discussion of the dualization in the introduction, we
have defined the pseudoscalar axion as $\partial_\sigma a = \sqrt{2}
\epsilon_{\mu\nu\rho\sigma}H^{\mu\nu\rho}$, or in form notation,
\begin{equation}
d\tilde a = \frac{\sqrt{2}}{2\kappa}~{^*{\tilde H}}
\label{A19}
\end{equation}
\noindent We skip the laborious algebra leading towards the final answer, and
only remark that the inclusion of the gauge Chern-Simons form in the definition
proves crucial for integrability of $d\tilde a$. The final expression for
$\tilde a$ is
\begin{equation}
\tilde a = -\frac{\sqrt{2}\alpha'}{16\kappa}\frac{Q^2_e
A}{GM}\frac{\cos\theta}{\tilde \rho^2 + A^2 \cos\theta}
\end{equation}
\noindent which to order $O(\alpha',A)$ is identical with our solution.
}
}
\vfill
\eject

\newpage

\noindent{\bf{Figure Captions}}

\vskip.3truein

\begin{itemize}

\item[]
\begin{enumerate}
\item[]
\begin{enumerate}
\item[{\bf Figure 1:}] The dilaton a) and axion b) solutions of
CKO (solid curve) and GMGHSSTWS
(dashed curve) as a function of the radial coordinate $r$ (see text).
Parameters chosen
are $G M^2 = 1$ and $Q'^2 = \frac{1}{8}$
with $\alpha' = 1$. In b) we have taken $A/GM = 0.1$

\item[{\bf Figure 2:}] Same as Figure 1 with $G M^2 = 10^4$ and
$(Q'^2/G M^2) = 1.25 \times 10^{-3}$,  and in b)  $A/GM = 0.1$.

\item[{\bf  Figure 3:}] Same as Figure 1 with $G M^2 = 10^4$
 and $(Q'^2/G M^2) = 1.53$, and in b) we have taken $A/GM = 0.5$.
\end{enumerate}
\end{enumerate}
\end{itemize}


\newpage

\begin{thebibliography}{99}
\vskip.3truein

\bibitem{strev} M. Green, J. Schwarz, and E. Witten,
$\underline{Superstring ~Theory}$, Cambridge University Press (1987); D. Lust
and S. Theisen, $\underline{Lectures ~on ~String ~Theory}$ Springer-Verlag
(1990).

\bibitem{2}C. Lovelace, Phys. Lett. {\bf B135} (1984) 75;
 E.S. Fradkin and A.A. Tseytlin, Phys. Lett. {\bf B158} (1985) 316;
 C.G. Callan, D. Freidan, E.J. Martinec, and M.J. Perry,
 Nucl. Phys. {\bf B262} (1985) 593;
 A. Sen, Phys. Rev. {\bf D32} (1985) 2102;
 Phys. Rev. Lett. {\bf 55} (1985) 1846;
 C.G. Callan, I.R. Klebanov, and M.J. Perry, Nucl. Phys. {\bf B278} (1986) 78;
 S. Deser and N. Redlich, Phys. Lett. {\bf B176} (1986) 350;
 D. Luest, S. Theisen and G. Zoupanos, Nucl. Phys. {\bf B296} (1988) 800;
 J. Lauer, D. Luest, and S. Theisen, Nucl. Phys. {\bf B304} (1988) 236.

\bibitem{3} D. Gross and J. Sloan, Nucl. Phys. {\bf B291} (1987) 41.

\bibitem{CMP} C. Callan, R. Myers and M. Perry,
Nucl. Phys. {\bf B251} (1989) 34.

\bibitem{CDKO1} B.A. Campbell, M.J. Duncan, N. Kaloper and K.A. Olive,
Nucl. Phys. {\bf B351} (1991) 778.

\bibitem{CDKO2} B.A. Campbell, M.J. Duncan, N. Kaloper and K.A. Olive,
Phys. Lett. {\bf B251} (1990) 34.

\bibitem{CKO1} B.A. Campbell, N. Kaloper and K.A. Olive,
Phys. Lett. {\bf B263} (1991) 364.

\bibitem{LW}  K.-Y. Lee and E. Weinberg, Phys. Rev {\bf D44} (1991) 3159.

\bibitem{Reut} M. Reuter, Class.Quant.Grav. {\bf 9} (1992) 751.

\bibitem{CKO2} B.A. Campbell, N. Kaloper and K.A. Olive,
Phys. Lett. {\bf B285} (1992) 199.

\bibitem{Gibb} G.W. Gibbons, Nucl. Phys. {\bf B207} (1982) 337.

\bibitem{GM} G.W. Gibbons and K. Maeda, Nucl. Phys. {\bf B298} (1988) 741.

\bibitem{GHS} D. Garfinkle, G. Horowitz and A. Strominger,
Phys. Rev. {\bf D43} (1991) 3140.

\bibitem{STW} A. Shapere, S. Trivedi and F. Wilczek,
Mod. Phys. Lett. {\bf A6} (1991) 2677.

\bibitem{PSSTW} J. Preskill, P. Schwarz, A. Shapere, S. Trivedi and F. Wilczek,
 Mod. Phys. Lett. {\bf A6} (1991) 2353.

\bibitem{HW} C. Holzhey and F. Wilczek,  Nucl. Phys.
{\bf B380} (1992) 447.

\bibitem{CPW1} S. Coleman, J. Preskill and F. Wilczek, Phys. Rev. Lett.
{\bf 67} (1991) 2677.

\bibitem{CPW2} S. Coleman, J. Preskill and F. Wilczek,  Nucl. Phys.
{\bf B378} (1992) 1975.

\bibitem{Sen} A. Sen, Phys. Rev. Lett {\bf 69} (1992) 1006.

\bibitem{kal1} R. Kallosh, Phys. Lett. {\bf B282} (1992) 80.

\bibitem{kal2} R. Kallosh, A. Linde, T. Ort\'in, A. Peet and A. Van Proyen,
Stanford Univ. preprint SU-ITP-92-13, May 1992.

\bibitem{kal3} R. Kallosh and A. Peet, Stanford Univ. preprint SU-ITP-92-27,
Sep. 1992.

\bibitem{kal4} R. Kallosh, T. Ort\'in and A. Peet, Stanford Univ. preprint
SU-ITP-92-29, Nov. 1992.

\bibitem{BBJ} M. Bento and O. Bertolami, Phys. Lett. {\bf B228} (1989) 348;
Phys. Lett. {\bf B220} (1989) 113; Phys. Lett. {\bf B218} (1989) 162;
I. Jack, D.R.T. Jones and N. Mohammedi, Phys. Lett. {\bf B220} (1989) 176.

\bibitem{DKO} M.J. Duncan, N. Kaloper and K.A. Olive, Univ. of Minnesota
preprint UMN-TH-1102/92, Feb. 1992, in press in Nucl. Phys. {\bf B}.

\bibitem{CFJ} S.M. Carrol, G. Fields and R. Jackiw,
Phys. Rev. {\bf D41} (1990) 1231; S.M. Carrol and G. Fields,
Phys. Rev. {\bf D43} (1991) 3789

\end{thebibliography}
\end{document}